\documentclass[prl,superscriptaddress,twocolumn,epsf]{revtex4}
\usepackage{epsfig}

\def \binom#1 #2{{#1\choose #2}}

\begin{document}
\title{Nuclear Spins in a Nanoscale Device for Quantum Information Processing}

\author{\c{S}. K. \"Ozdemir}
\affiliation{SORST-JST, Honmachi, Kawaguchi, 331-0012 Saitama,
Japan} \affiliation{CREST-JST, Honmachi, Kawaguchi, 331-0012
Saitama Japan} \affiliation{Graduate School of Engineering
Science, Osaka University, 1-3 Machikaneyama, Toyonaka, 560-8531
Osaka, Japan}

\author{A. Miranowicz}
\affiliation{SORST-JST, Honmachi, Kawaguchi, 331-0012 Saitama,
Japan} \affiliation{Graduate School of Engineering Science, Osaka
University, 1-3 Machikaneyama, Toyonaka, 560-8531 Osaka,
Japan}\affiliation{Institute of Physics, Adam Mickiewicz
University, 61-614 Pozna$\acute{n}$, Poland}

\author{T. Ota}
\affiliation{SORST-JST, Honmachi, Kawaguchi, 331-0012 Saitama,
Japan}\affiliation{NTT Basic Research Labs, NTT Corporation,
Atsugi, 243-0198 Kanagawa, Japan}

\author{G. Yusa}
\affiliation{NTT Basic Research Labs, NTT Corporation, Atsugi,
243-0198 Kanagawa, Japan}\affiliation{PRESTO-JST, Honmachi,
Kawaguchi, 331-0012 Saitama, Japan}\affiliation{Department of
Physics, Tohoku University, Sendai, 980-8578 Miyagi, Japan}

\author{N. Imoto}
\affiliation{SORST-JST, Honmachi, Kawaguchi, 331-0012 Saitama,
Japan} \affiliation{CREST-JST, Honmachi, Kawaguchi, 331-0012
Saitama Japan} \affiliation{Graduate School of Engineering
Science, Osaka University, 1-3 Machikaneyama, Toyonaka, 560-8531
Osaka, Japan}

\author{Y. Hirayama}
\affiliation{SORST-JST, Honmachi, Kawaguchi, 331-0012 Saitama,
Japan} \affiliation{Department of Physics, Tohoku University,
Sendai, 980-8578 Miyagi, Japan}

\begin{abstract}
Coherent oscillations between any two levels from four nuclear
spin states of $I=3/2$ have been demonstrated in a nanometre-scale
NMR semiconductor device, where nuclear spins are all-electrically
controlled. Using this device, we discuss quantum logic operations
on two fictitious qubits of the $I=3/2$ system, and propose a
quantum state tomography scheme based on the measurement of
longitudinal magnetization, $M_z$.
\end{abstract}
\pagestyle{plain} \pagenumbering{arabic} \maketitle

\section {Introduction}

It has long been demonstrated that nuclear spins have long
relaxation times, thus longer coherence times,  making them
suitable for quantum information processing devices
\cite{Cory1,Gershenfeld} (for a review see \cite{Vandersypen} and
references therein). Most of the works on the manipulation and use
of nuclear spins are restricted to those in molecules of liquid
solutions. To date, liquid-state NMR is the leader among all
quantum computer implementations, both in the number of qubits
controlled and the number of gates performed within the coherence
lifetime. These successful demonstrations inspired researches to
investigate the possibility to use NMR in solid-state systems for
the same purpose \cite{Cory}. Coherent manipulation of nuclear
spins in solid-state systems became an exciting field of research
which resulted in interesting proposals on nuclear spin qubits,
where nuclear spins are individually manipulated \cite{Kane,Ladd}.

An advantage of solid-state NMR over the liquid NMR is that the
spins can be highly polarized by dynamic nuclear spin polarization
techniques such as polarization transfer from electronic spins.
This strong polarization allows a large population available for
the preparation of the system close to a pure state. However, only
a few quantum information processing experiments with
single-crystal solids have been reported \cite{Cory,Kamper}.

Most of the experiments up to date employ conventional NMR
technique, which uses metal coils for excitation and detection of
nuclear spin transverse-magnetization $M_{xy}$. In order to obtain
a detectable signal, a large number of nuclear spins
($10^{11}-10^{13}$) should be involved in the process, which
limits the sensitivity of these schemes. Moreover, quantum
multiple coherence is not directly detected in conventional NMR as
it does not produce transverse magnetization.

In a recent study \cite{Yusa}, Yusa {\em et al.} demonstrated an
all-electrical control and detection of nuclear spin magnetization
in a nanometre-scale semiconductor device, where longitudinal
magnetization $M_z$ of nuclear spins in a point-contact channel
are detected by the resistivity change of the channel paving the
way to the detection of multiple quantum coherence. This device
has an improved sensitivity, which requires the involvement of
only $\sim 10^8$ or less nuclear spins.

In the following, we first describe this nanoscale NMR device, the
novel detection technique and the observed multiple quantum
coherence. Then, we show how one- and two-qubit operations can be
performed on this device, and give the pulse sequences. A method
for performing quantum state tomography using $M_z$-detection in
this device is also presented, and finally we give a brief
discussion and conclusion.

\section {Nanoscale Semiconductor NMR device}

{\bf Principle of the device:} The nanoscale NMR device is
fabricated as a monolithic semiconductor device integrated with a
point contact channel and an antenna gate (see Fig. 1). The
structure contains a 20-nm GaAs quantum well with AlGaAs barrier
layers grown on n-GaAs(100) substrate. This substrate functions as
backgate to control the electron density in the point contact
region, which is defined by split Schottky gates separated by
600nm. The antenna gate locally irradiates the channel with an
alternating rf field for selective and coherent manipulation of
the nuclear spins.

By controlling the static magnetic field perpendicular to the
grown device surface and the backgate voltage, the point contact
region was set to fractional quantum Hall regime at the degenerate
Landau-level filling factor $\nu=2/3$, where there is a strong
coupling of nuclear spins to the conduction electrons
\cite{Kron,Hashimoto,Smet}. At $\nu=2/3$, the spin polarized and
unpolarized states face each other through a transition region
with high resistivity $R_{xx}$. When sufficient current is driven
through the system, the polarized nuclear spins interact with the
flowing electrons resulting in an gradual enhancement of $R_{xx}$.
Thus, the current density becomes very high in the constricted
region so that nuclear spin polarization occurs only in the point
contact region. This enhancement of $R_{xx}$ implies that the
polarization of the nuclear spins in the point contact region can
be detected by measuring the resistance between the ends of the
point contact. It has already been shown that this resistance
values is proportional to the longitudinal magnetization of
nuclear spins, $\Delta M_z \propto \Delta R_{xx}$
\cite{Hashimoto}.
\begin{figure}
\epsfxsize=7cm \epsfbox{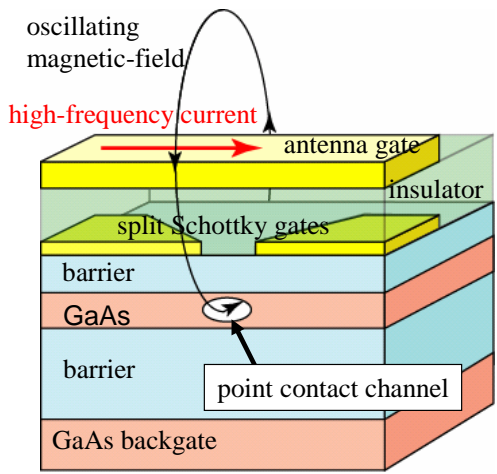} \caption{Schematic diagram
illustrating the semiconductor device for coherent-control of
nuclear spins in a nanometre-scale region. The point contact
channel, which is pointed with an arrow, is defined by a Schottky
gate pair.} \label{fig1}
\end{figure}

{\bf Observation of coherent oscillation with  $R_{xx}$
measurement:} The point contact channel in this device consists of
$^{69,71}{\rm Ga}$ and $^{75}{\rm As}$ isotopes each having total
spin $I=3/2$. Thus, each nuclide splits into $n=2I+1=4$ energy
states,
$|m\rangle=|3/2\rangle,|1/2\rangle,|-1/2\rangle,|-3/2\rangle$,
under static magnetic field due to Zeeman effect. This levels are
spaced with equal energy separation of $\hbar \omega_0$. However,
the quadrupolar interaction shifts the adjacent states by
$2\Delta_q$ from $\hbar \omega_0$ allowing three possible
transitions at resonances $\hbar \omega_0-2\Delta_q$, $\hbar
\omega_0$ and $\hbar \omega_0+2\Delta_q$ (see Fig. 3). While these
transitions are between levels separated by one quantum angular
momentum ($\Delta m=1$), energy and angular momentum conservation
rules imply that transitions between levels separated by $\Delta
m=2$ and $\Delta m=3$ are also possible, resulting in a total of
six possible coherent oscillations. We named those transitions as
one-photon, two-photon and three-photon transitions. Therefore, in
this NMR device it is possible to observe six coherent transitions
for each of the three nuclei ($^{69,71}{\rm Ga}$ and $^{75}{\rm
As}$) in the point contact channel.

In order to observe the coherent oscillations, first the nuclear
spin polarization is saturated and this is observed by resistance
measurement, which is also saturated. Then pulsed rf magnetic
field is applied by the antenna gate. If the frequency of the
applied pulse is in resonance with the NMR frequency, oscillatory
change in $M_z$ is observed implying that the superposition
coherently rotates between the two energy states. The oscillations
in $M_z$ is finally detected by the change in the resistance,
$\Delta R_{xx}$, before and after the application of the rf pulse.
In Fig.2, NMR spectrum for one- and two- photon transitions are
given only for $^{75}{\rm As}$. Coherent oscillations are observed
for three-photon transitions, too (not shown here). In the same
way, all of the six possible oscillations are observed for
$^{69}{\rm Ga}$ and $^{71}{\rm Ga}$ \cite{Yusa}.

The decoherence time $T_2$ of the device is estimated by curve
fitting the coherent oscillation between $|-1/2\rangle$ and
$|-3/2\rangle$ of $^{75}As$. The $T_2$ was estimated as $\sim 0.6
ms$ without any decoupling process. However, it is enhanced to
$1.5 ms$ when the nuclei-electron decoupling is applied
\cite{Hirayama}. The relaxation time $T_1$ in this device is
longer than $100 s$.

\section {State manipulations in the nanoscale NMR device}

The four-level system in this device forms a quartit, which is
equivalent to two logical qubits. This becomes clear if we
identify $|0\rangle\equiv|00\rangle_{AB}\equiv|3/2\rangle$,
$|1\rangle\equiv|01\rangle_{AB}\equiv|1/2\rangle$,
$|2\rangle\equiv|10\rangle_{AB}\equiv|-1/2\rangle$, and
$|3\rangle\equiv|11\rangle_{AB}\equiv|-3/2\rangle$ (see Fig. 2).
Thus, the two-qubit state can be written as
$|\psi\rangle=c_0|0\rangle+c_1|1\rangle+c_2|2\rangle+c_3|3\rangle$,
where $\sum_i |c_i|^2=1$. In the following, we show how different
quantum gate operations can be performed on this two fictitious
qubit system by applying selective pulses at the resonant
frequency between two energy levels, say $|m\rangle$ and
$|n\rangle$. Here, the free evolution of the coherent system
during the finite pulse duration and the time lag between the
pulses are ignored, however they can be compensated during
experimental realizations.

\begin{figure}
\epsfxsize=8cm \epsfbox{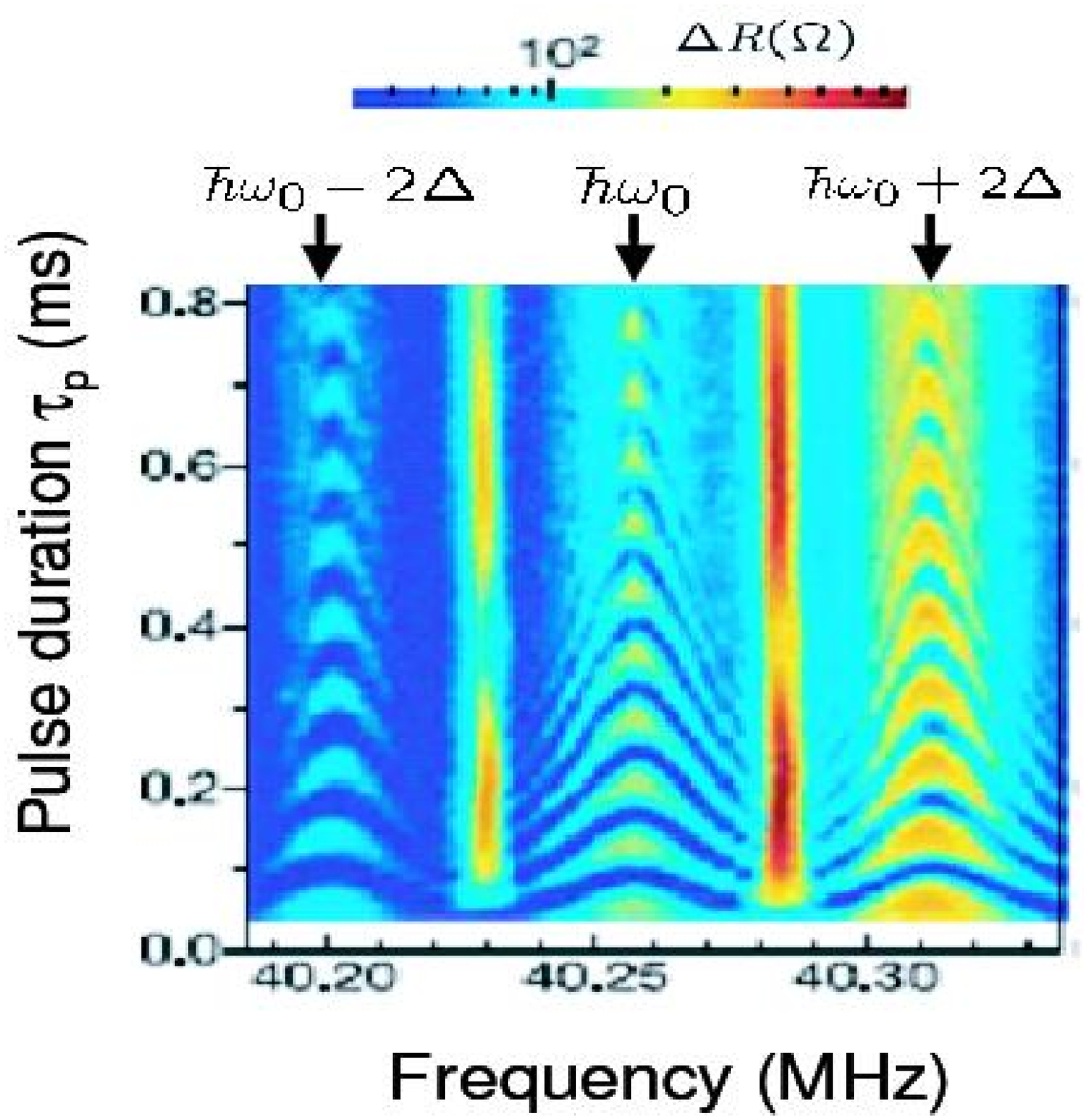}

\vspace*{4mm} \epsfxsize=8cm \epsfbox{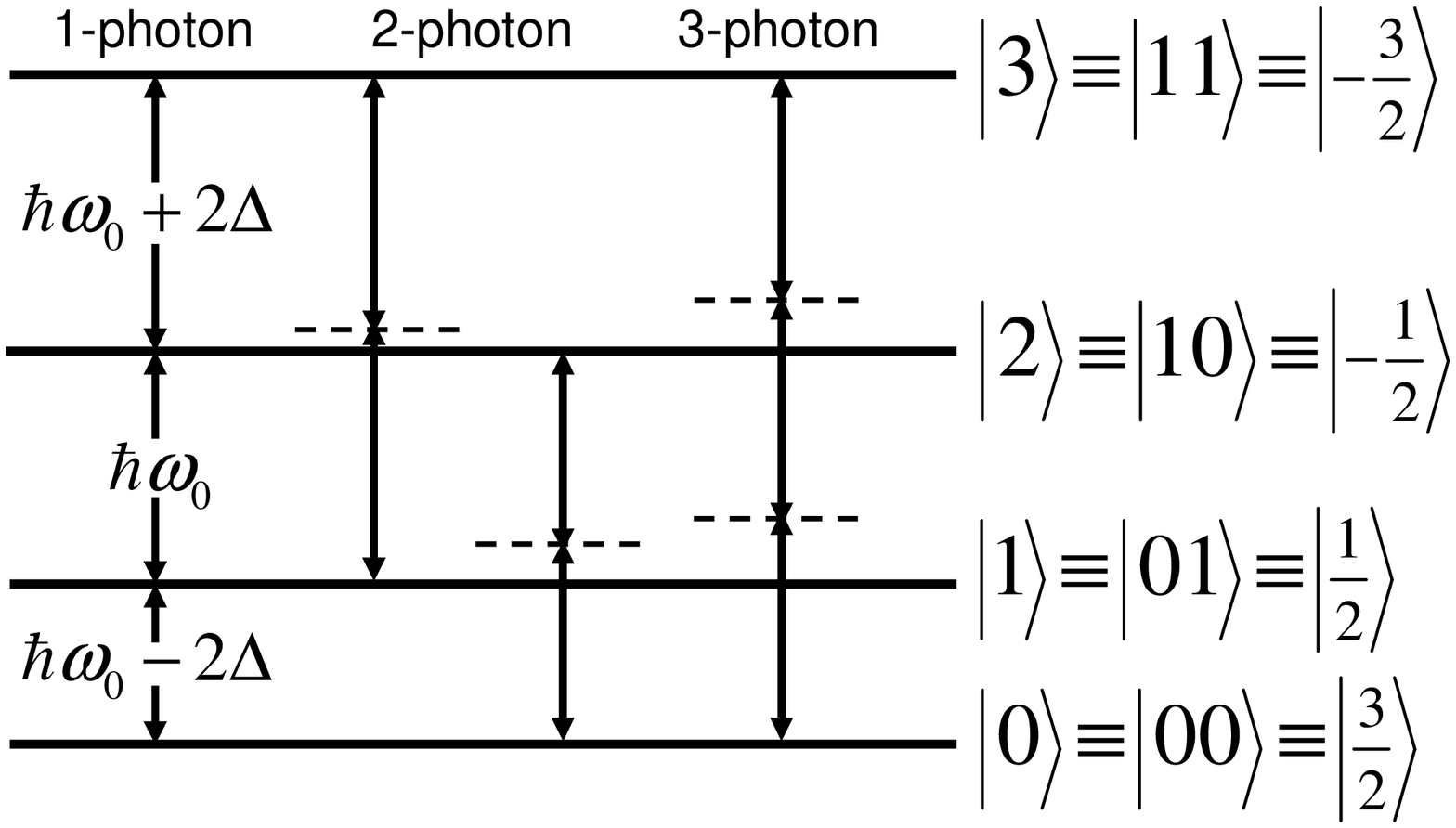}
\caption{Schematic energy level diagram of nuclear spin states for
I=3/2 with electric quadrupolar interactions (lower figure) and
coherent oscillations corresponding to single-photon  and
two-photon transitions between the four spin states observed for
$^{75}{\rm As}$ (upper). A three-photon transition was also
observed for higher values of the alternating current through the
antenna gate (data not shown). Measurements were performed at 100
mK with a magnetic field of 5.5T.} \label{Fig2}
\end{figure}

Rotations of logical qubit A in this quartit can be obtained by
applying rf pulses at frequencies $\omega_0-\Delta_q/\hbar$ and
$\omega_0+\Delta_q/\hbar$ two induce two-photon transitions:
$\hat{X}_A(\theta)=\hat{X}_{02}(\theta)\hat{X}_{13}(\theta)$,
$\hat{Y}_A(\theta)=\hat{Y}_{02}(\theta)\hat{Y}_{13}(\theta)$, and
$\hat{Z}_A(\theta)=\hat{Z}_{02}(\theta)\hat{Z}_{13}(\theta)$,
where $\hat{X}_{nm}(\theta), \hat{Y}_{nm}(\theta)$ and
$\hat{Z}_{nm}(\theta)$ corresponds to rotations between levels
$|m\rangle$ and $|n\rangle$ by an angle $\theta$ along the
corresponding axes. In the same way, rotations on the logical
qubit B can be obtained by applying pulses at frequencies
$\omega_0-2\Delta_q/\hbar$ and $\omega_0+2\Delta_q/\hbar$ two
induce one-photon transitions:
$\hat{X}_B(\theta)=\hat{X}_{01}(\theta)\hat{X}_{23}(\theta)$,
$\hat{Y}_B(\theta)=\hat{Y}_{01}(\theta)\hat{Y}_{23}(\theta)$, and
$\hat{Z}_B(\theta)=\hat{Z}_{01}(\theta)\hat{Z}_{23}(\theta)$.

{\bf NOT Gate:} The transformation of this two-qubit system when a
NOT-gate is applied to the first (A) and second (B) logical qubits
is described by $ \hat{U}_{\rm NOT}^{A}|\psi\rangle=
c_0|2\rangle+c_1|3\rangle+c_2|0\rangle+c_3|1\rangle $ and $
\hat{U}_{\rm NOT}^{B}|\psi\rangle=
c_0|1\rangle+c_1|0\rangle+c_2|3\rangle+c_3|2\rangle $, which can
be implemented by the selective pulse sequences $\hat{U}_{\rm
NOT}^{A}=i\hat{X}_{02}(\pi)\hat{X}_{13}(\pi)$ and $\hat{U}_{\rm
NOT}^{B}=i\hat{X}_{01}(\pi)\hat{X}_{23}(\pi)$. On the other hand,
NOT-gate on both qubits can be obtained by a $\pi$-hard-pulse
resulting in $\hat{U}_{\rm NOT}^{AB}|\psi\rangle=
c_0|3\rangle+c_1|2\rangle+c_2|1\rangle+c_3|0\rangle$. One can
easily see that $\hat{U}_{\rm NOT}^{A}$, which is implemented by
pulses at two-photon transitions can be implemented by a
hard-pulse followed by $\hat{U}_{\rm NOT}^{B}$.

{\bf Hadamard Gate:} This truly quantum gate transforms the
computational basis states into an equally weighted superposition
states. A Hadamard operation on the first qubit makes the
transformation $ \hat{U}_{\rm H}^{A}|\psi\rangle=
(c_0+c_2)|0\rangle+(c_1+c_3)|1\rangle+(c_0-c_2)|2\rangle+(c_1-c_3)|3\rangle
$ and can be implemented by the pulse sequence $ \hat{U}_{\rm
H}^{A}=i\hat{Y}_{12}(\pi)\hat{X}_{01}(\pi)\hat{Y}_{01}(\pi/2)\hat{X}_{23}(-\pi)\hat{Y}_{23}(-\pi/2)\hat{Y}_{12}(-\pi)$.
In the same way, Hadamard on the second qubit transforms the
initial state as $ \hat{U}_{\rm H}^{B}|\psi\rangle=
(c_0+c_1)|0\rangle+(c_0-c_1)|1\rangle+(c_2+c_3)|2\rangle+(c_2-c_3)|3\rangle
$ with the pulse sequence $ \hat{U}_{\rm
H}^{B}=i\hat{X}_{01}(\pi)\hat{Y}_{01}(\pi/2)\hat{X}_{23}(\pi)\hat{Y}_{23}(\pi/2)$.

{\bf CNOT Gate:} In order to realize quantum algorithms on this
NMR device, we should also show that effective realization of
two-qubit operations are possible in this system. A CNOT gate with
the first qubit as the control qubit and the second one as the
target qubit is defined by the transformation $\hat{U}_{\rm
CNOT}^{AB}|\psi\rangle=
c_0|0\rangle+c_1|1\rangle+c_3|2\rangle+c_2|3\rangle$. On the other
hand, the CNOT with the first qubit as the target and the second
qubit as the control performs the transformation $\hat{U}_{\rm
CNOT}^{BA}|\psi\rangle=
c_0|0\rangle+c_3|1\rangle+c_2|2\rangle+c_1|3\rangle$. One can
easily see that in this NMR device CNOT-like transformations can
be implemented by just one pulse as $\hat{U'}_{\rm
CNOT}^{AB}=\hat{Y}_{23}(\pi)$ and $\hat{U'}_{\rm
CNOT}^{AB}=\hat{Y}_{13}(\pi)$. These gates differ from the ideal
CNOT gate by an extra minus sign in one of the off-diagonal terms
of the transformation matrix of the corresponding gate operation,
i.e., $\hat{U'}_{\rm CNOT}^{AB}|\psi\rangle=
c_0|0\rangle+c_1|1\rangle-c_3|2\rangle+c_2|3\rangle$.  One can
alternatively perform the same operations by applying
$\hat{U}_{\rm CNOT}^{AB}=\hat{X}_{23}(\pi)$ and $\hat{U}_{\rm
CNOT}^{AB}=\hat{X}_{13}(\pi)$. On the other hand, an ideal CNOT
gate can be implemented by allowing a more complex sequence of
one-photon transition pulses. As an example of this, we give the
pulse sequence for $\hat{U}_{\rm CNOT}^{AB}$, which is
$\hat{U}_{\rm CNOT}^{AB}=\hat{U}_{\rm
H}^{B}\hat{Z}_{23}(\pi)\hat{Y}_{12}(-\pi)\hat{Z}_{23}
(\pi/2)\hat{Z}_{01}(\pi/2)\hat{Y}_{12}(\pi)\hat{U}_{\rm H}^{B}$,
where $\hat{U}_{\rm H}^{B}$ is defined as above. One may further
simplify this pulse sequence by using combination of one- and
two-photon transition pulses.

{\bf SWAP Gate:} In principle, a swap gate, which employs the
transformation $\hat{U}_{\rm swap}^{AB}|\psi\rangle=
c_0|0\rangle+c_1|2\rangle+c_2|1\rangle+c_3|3\rangle$, can be
implemented by a sequence of CNOT operations as follows
$\hat{U}_{\rm swap}^{AB}=\hat{U}_{\rm CNOT}^{AB}\hat{U}_{\rm
CNOT}^{BA}\hat{U}_{\rm CNOT}^{AB}$. On the other hand, for this
device we can obtain SWAP-like gates by a single pulse
$\hat{U}_{\rm swap}^{AB}=\hat{Y}_{12}(\pi)$ or $\hat{U}_{\rm
swap}^{AB}=\hat{X}_{12}(\pi)$. This, too, differs from the ideal
operation with a minus sign, which can be compensated by applying
more complex pulse sequences.
\begin{figure*}
\epsfxsize=4cm \epsfbox{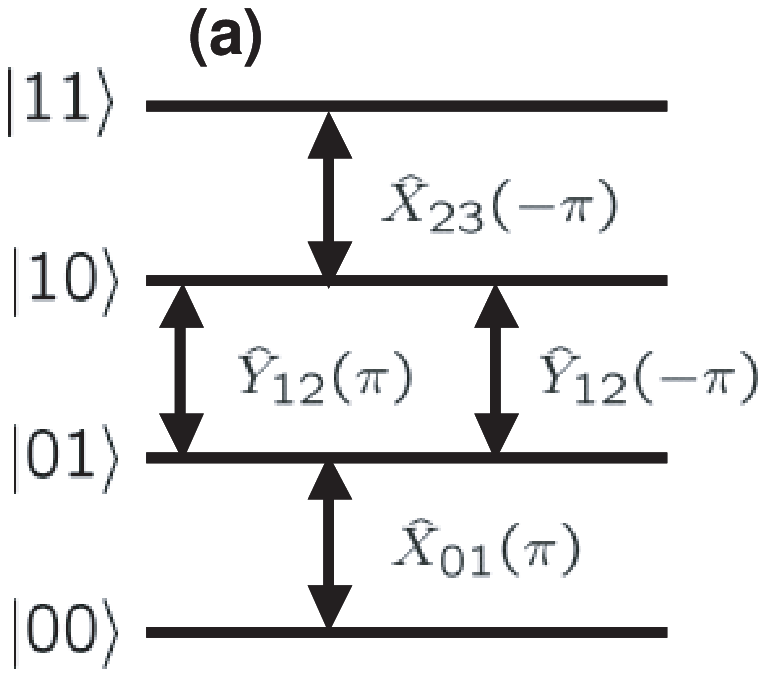} \hspace{10mm} \epsfxsize=3.5cm
\epsfbox{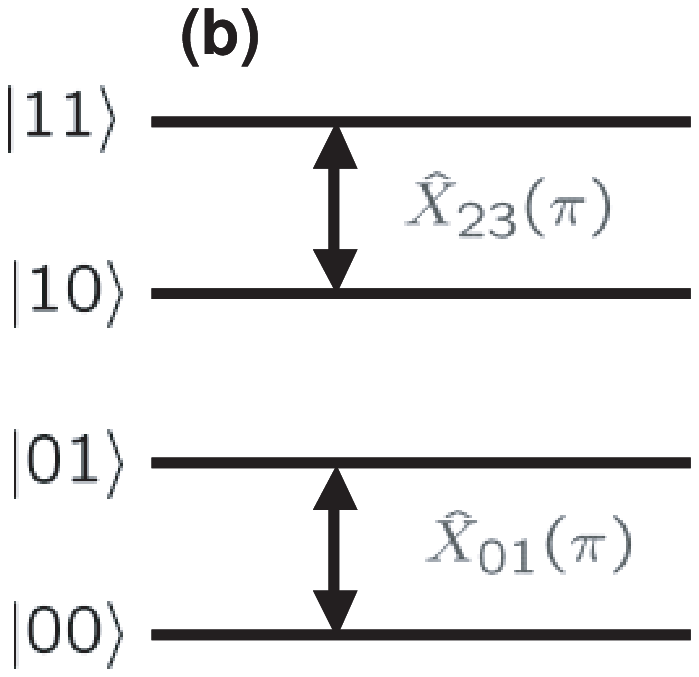}\hspace{10mm}\epsfxsize=4.1cm \epsfbox{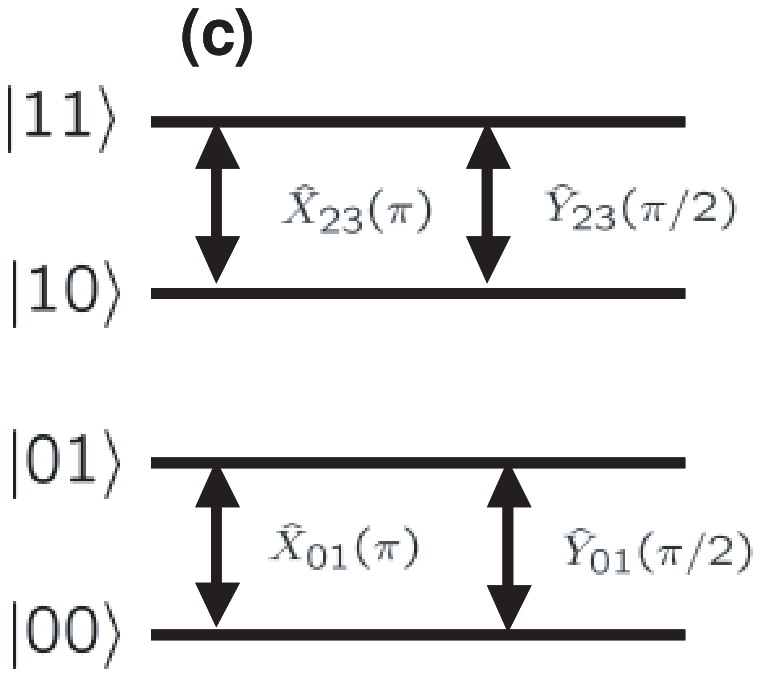}\\
\vspace{5mm} \epsfxsize=4.5cm
\epsfbox{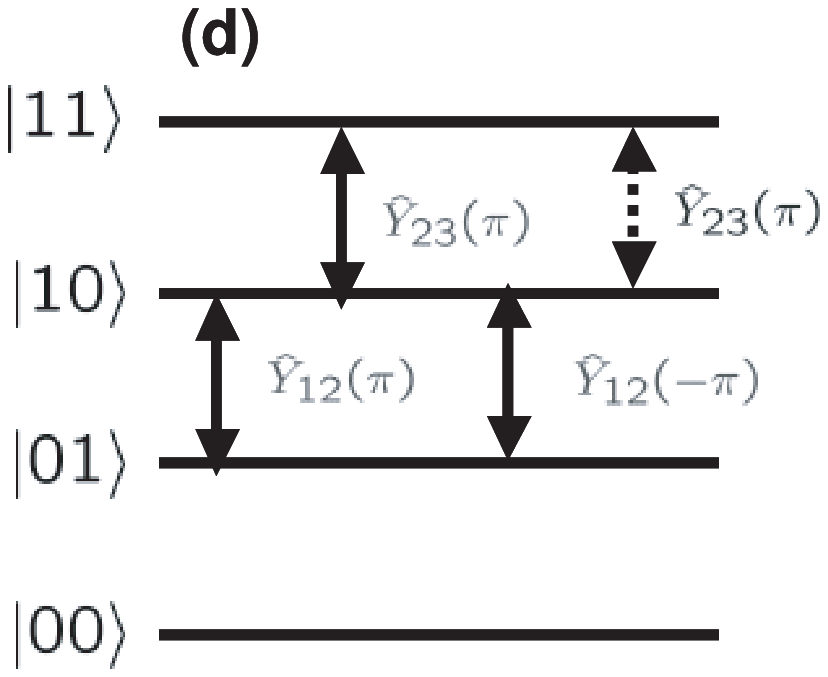}\hspace{10mm}\epsfxsize=6.7cm
\epsfbox{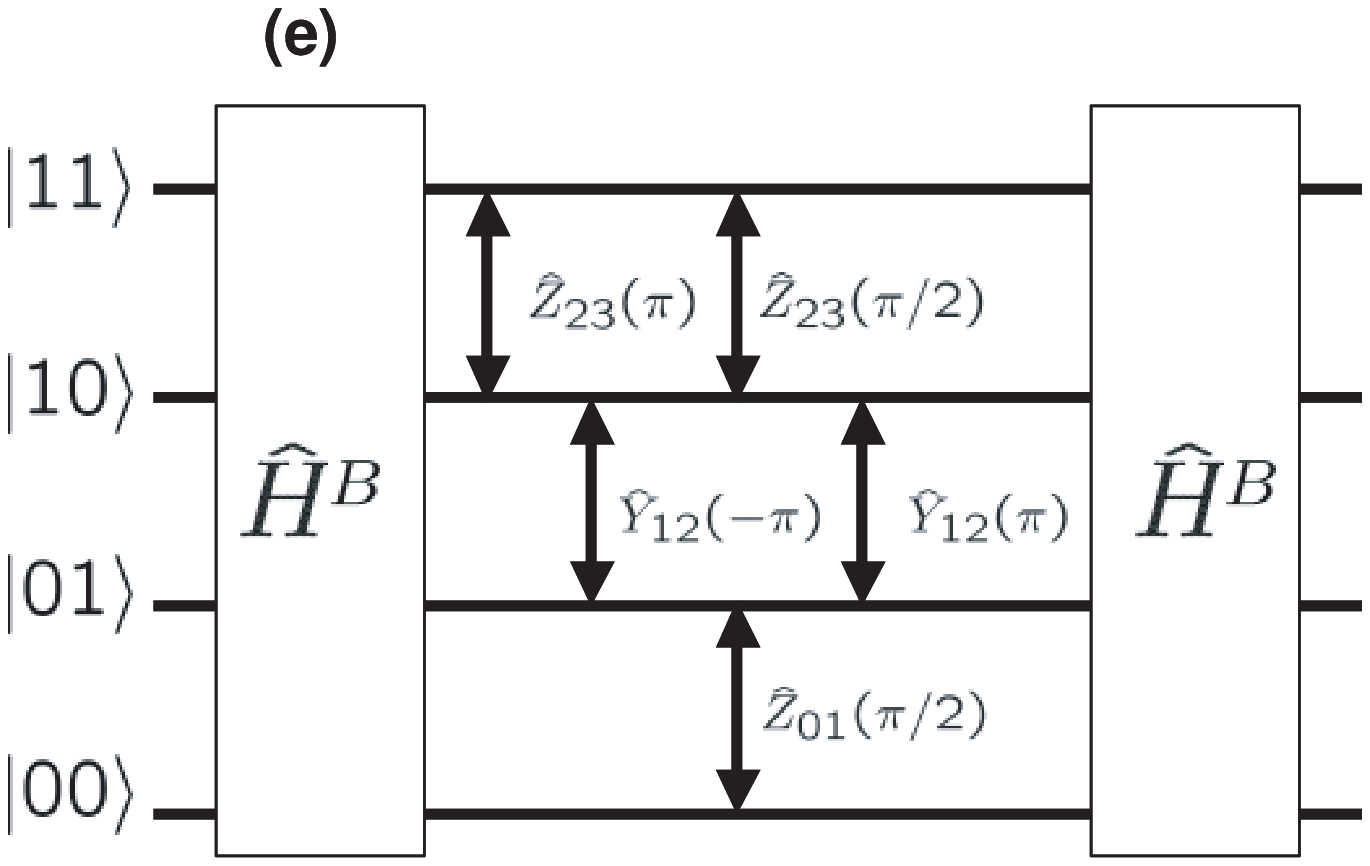}\caption{Illustration of pulse sequences to
realize (a) NOT gate on the first qubit $A$, (b) NOT gate on the
second qubit $B$, (c) Hadamard gate on the second qubit $B$, (d)
CNOT-like gate, and (e) ideal CNOT gate. In (d) the dotted
(straight) arrow corresponds to the case when first (second) qubit
is control and the second (first) qubit is target. In (e)
$\hat{H}^B$ is the Hadamard gate illustrated in (c).} \label{fig3}
\end{figure*}

\begin{figure*}
\epsfxsize=14cm \epsfbox{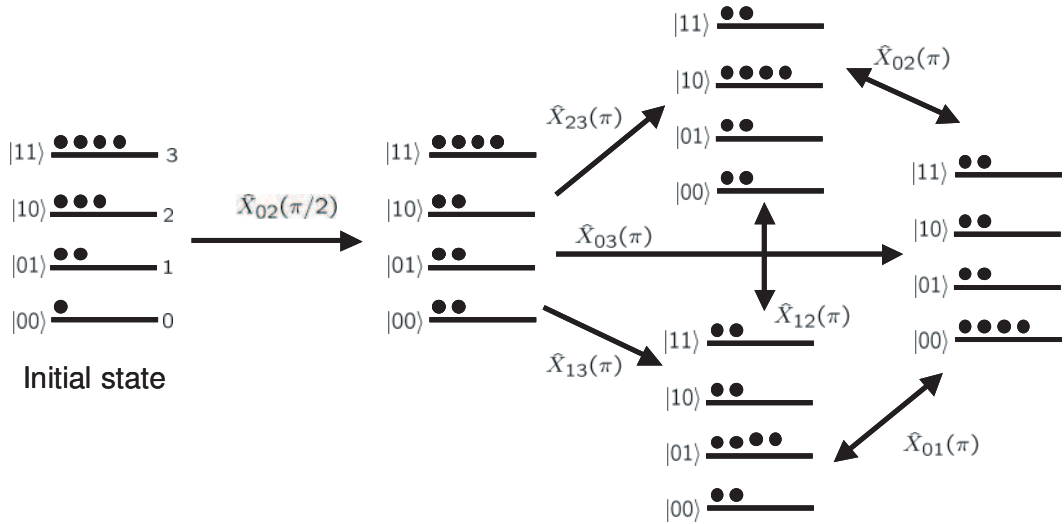}\caption{A schematic
illustration of pulse sequences for effective pure state
preparation starting from the initial state of the semiconductor
NMR device. Note that two-photon and three-photon transitions can
be decomposed into one-photon transitions using more complicated
pulse sequences, i.e, $X_{02}(\pi/2)=X_{01}(\pi/2)X_{12}(\pi)$.}
\label{fig4}
\end{figure*}

{\bf Effective pure state preparation:} For the correct working of
the gate operations and consequently the quantum computer, NMR
quantum computer should be properly prepared in an effective pure
initial state of the nucleus ensemble. Contrary the conventional
NMR, in our semiconductor NMR device we can prepare a large
population available as the pure state. We owe this to the
strongly polarized situation created due to dynamic nuclear spin
polarization induced by the current flow. The initial state of the
device is far from a pure state as schematically shown in
\ref{fig4}. Note that due to dynamic polarization the population
becomes larger for the higher spin states. From this initial
state, one can apply $\hat{X}_{01}(\pi/2)\hat{X}_{12}(\pi)$ to
prepare the $|11\rangle$. Once this state is prepared then we can
prepare the pure state $|ij\rangle$ by applying the pulse
$\hat{X}_{2i+j,3}(\pi)$ (see Fig. 4). More complicated pulse
sequences allows us to prepare effective pure state from any
arbitrary distribution. The designed population can be maintained
up to $T_1$, which is longer than $100s$ for this device.
Experimental results showing effective pure state preparation can
be found in Ref. \cite{Hirayama}.

\section {Quantum-state tomography based on $M_z$ detection}

Quantum state tomography (QST) is a method for complete
reconstruction of a given density matrix $\hat{\rho}$ in a serious
of measurements. In general, to reconstruct completely a density
matrix $\hat{\rho}$ for a quartit or two qubits, we need to
determine 15 real parameters (The 16th element can be found from
the normalization condition). Single NMR read-out can only give
some of the either diagonal or off-diagonal elements of the given
density matrix $\hat{\rho}$.

In case of the conventional NMR systems, where an $M_{xy}$
measurement is performed, a single measurement gives directly some
of the off-diagonal elements of the density matrix. However, in
our NMR device, where an $M_z$ measurement is employed, a single
measurement determines only the population differences
$\rho_{nn}-\rho_{mm}$ between the levels $|n\rangle$ and
$|m\rangle$. This quantity is related to the diagonal elements of
the density matrix, where $\rho_{mn}\equiv \langle
m|\hat{\rho}|n\rangle$. In a typical NMR spectrum of our device
peaks located at $\omega_{01}$, $\omega_{12}$, and $\omega_{23}$
give, respectively, $\rho_{11}-\rho_{00}$, $\rho_{22}-\rho_{11}$,
and $\rho_{33}-\rho_{22}$. By imposing the normalization
condition, we can easily obtain the diagonal elements $\rho_{ii}$
of the density matrix. The remaining elements of the density
matrix can be obtained by rotating the original density matrix by
properly chosen rotation operations $\hat{R}_k$, which transform
the original matrix into $\hat{\rho}_k=\hat{R}_k \rho
\hat{R}_k^{\dagger}$.  These rotations move the off-diagonal
elements of the original density matrix to the diagonal of the
rotated density matrix so that $M_z$ measurement provides
information on them. By application of a number of such rotations,
the off-diagonal elements of the density matrix can be brought
into the measurable ones in $M_z$ detection. The set of
$\hat{R}_k$ operations is not unique, one can find many different
sets of rotations for complete reconstruction of the density
matrix using $M_z$ detection.
We found that the following set of 12 rotations is suitable for
the reconstruction of the density matrix using $M_z$ detection
based state tomography for our device
\begin{eqnarray}
&&\hat{R}_1=\hat{X}_{01}(\pi/2), ~~~\hat{R}_2=\hat{Y}_{01}(\pi/2), ~~~\hat{R}_3=\hat{X}_{12}(\pi/2),\nonumber \\
&&\hat{R}_4=\hat{Y}_{12}(\pi/2), ~~~\hat{R}_5=\hat{X}_{23}(\pi/2), ~~~\hat{R}_6=\hat{Y}_{23}(\pi/2),\nonumber \\
&&\hat{R}_7=\hat{X}_{02}(\pi/2), ~~~\hat{R}_8=\hat{Y}_{02}(\pi/2), ~~~\hat{R}_9=\hat{X}_{13}(\pi/2),\nonumber \\
&&\hat{R}_{10}=\hat{Y}_{13}(\pi/2), ~~\hat{R}_{11}=\hat{X}_{03}(\pi/2), ~~\hat{R}_{12}=\hat{Y}_{03}(\pi/2).\nonumber\\
\end{eqnarray}
One can optimize the rotation set to decrease the number of
rotations and/or increase the reconstruction sensitivity. The
following is one of such a set composed only of six operations and
still exhibiting high scheme sensitivity parameter. The rotations
in this set are \begin{eqnarray}
&&\hat{R}_1=\hat{X}_{01}(\pi/2)\hat{X}_{23}(\pi/2), \nonumber \\
&&\hat{R}_2=\hat{Y}_{01}(\pi/2)\hat{Y}_{23}(\pi/2), \nonumber \\
&&\hat{R}_3=\hat{X}_{01}(\pi/2)\hat{Y}_{13}(\pi)\hat{X}_{12}(\pi/2),\nonumber \\
&&\hat{R}_4=\hat{Y}_{01}(\pi/2)\hat{Y}_{13}(\pi)\hat{Y}_{12}(\pi/2),\nonumber \\
&&\hat{R}_5=\hat{X}_{02}(\pi/2)\hat{X}_{13}(\pi/2), \nonumber \\
&&\hat{R}_6=\hat{Y}_{02}(\pi/2)\hat{Y}_{13}(\pi/2)\nonumber\\
\end{eqnarray}

\section{Conclusions}

All-electrical coherent control of nuclear spins in a nanoscale
NMR chip has been demonstrated. Clear coherent oscillations
reflect all possible transitions among the four nuclear spin
states of each of the nuclide ($^{69,71}{\rm Ga}$ and $^{75}{\rm
As}$) in the point contact channel. Since arbitrary control of
superpositions among the four spin levels can be performed in this
device, one and two qubit operations are possible. We have
theoretically shown how to utilize this four-level system as a
fictitious two-qubit system, and designed the pulse sequences for
the realization of one- and two-qubit quantum operations.
Moreover, we proposed an NMR state tomography scheme based on the
detection of longitudinal magnetization of the nuclear spins.

Although we have not shown in this manuscript, we performed
preliminary experiments \cite{Hirayama} for the preparation of
effective pure states, which are important for the initialization
of the NMR quantum computer. Dynamic nuclear spin polarization
induced by the current flow allowed us to prepare a large
population to use for the preparation of effective pure states.

The results shown in this paper suggest that this nanometre-scale
NMR device is a good candidate for quantum-information processing
based on solid-state systems.

\noindent {\bf Acknowledgments}. \c{S}KO and AM kindly acknowledge
support from the Japan Society for the Promotion of Science within
the 21st Century COE Program. AM also acknowledges grant No. 1
P03B 064 28 of the Polish KBN.



\begin{thebibliography}{\mode}

\bibitem{Cory1}
D. G. Cory, A. F. Fahmy and T. F. Havel, PNAS USA {\bf 94}, 1634
(1997).

\bibitem{Gershenfeld}
N.~Gershenfeld and I.~L. Chuang, Science {\bf 275}, 350 (1997).

\bibitem{Vandersypen}
L. M. K. Vandersypen and I. L. Chuang,  Rev. Mod. Phys. {\bf 76},
1037 (2004).

\bibitem{Cory}
D. G. Cory {\it et al.}, Fortschr. Phys. {\bf 48}, 875 (2000).

\bibitem{Kane} B. E. Kane, Nature {\bf 393}, 133 (1999).

\bibitem{Ladd}
T. D. Ladd, J. R. Goldman, F. Yamaguchi, Y. Yamamoto, E. Abe, and
K. M. Itoh, Phys. Rev. Lett. {\bf 89}, 017901 (2002).

\bibitem{Kamper}
H. Kampermann, and W. S. Veeman, J. Chem. Phys. {\bf 122}, 214108
(2005).

\bibitem{Yusa}
G. Yusa, K. Muraki, K. Takashina, K. Hashimoto, and Y. Hirayama,
Nature {\bf 434}, 1001 (2005).

\bibitem{Kron}
S. Kronmuller, W. Dietsche, K. von Klitzing, G. Denninger, W.
Wegscheider, and M. Bichler, Phys. Rev. Lett. {\bf 82}, 4070
(1999).

\bibitem{Hashimoto}
K. Hashimoto, K. Muraki, T. Saku, and Y. Hirayama, Phys. Rev.
Lett. {\bf 88}, 176601 (2002).

\bibitem{Smet}
J. H. Smet, R. A. Deutschmann, F. Ertl, W. Wegscheider, G.
Abstreiter, and K. von Klitzing, Nature {\bf 415}, 281 (2002).

\bibitem{Hirayama}
Y. Hirayama, A. Miranowicz, T. Ota, G. Yusa, K. Muraki, \c{S}. K.
\"Ozdemir, and N. Imoto, J. Phys.: Condens. Matter {\bf 18}, S885
(2006).

\end{thebibliography}
\end{document}